\documentclass[aps,
10pt]{revtex4}
\usepackage{subfigure}
\usepackage{graphicx}
\usepackage{epsfig}
\usepackage{dcolumn}
\usepackage{bm}
\usepackage{times}
\usepackage{amsmath}
\usepackage{color}
\def\beq{\begin{equation}}
\def\eeq{\end{equation}}
\def\beqn{\begin{align}}
\def\eeqn{\end{align}}

\newcommand{\bs}{\bigskip}
\newcommand{\ms}{\medskip}

\begin{document}
\title{
Factorization conditions for nonlinear second-order differential equations 
}
\author{G. Gonz\'alez}
\email{gabriel.gonzalez@uaslp.mx}
\affiliation{C\'atedra CONCAYT--Universidad Aut\'onoma de San Luis Potos\'i, San Luis Potos\'i, 78000, Mexico}
\affiliation{Coordinaci\'on para la Innovaci\'on y la Aplicaci\'on de la Ciencia y la Tecnolog\'ia, Universidad Aut\'onoma de San Luis Potos\'i,San Luis Potos\'i, 78000, Mexico}
\author{H.C. Rosu}
\email{hcr@ipicyt.edu.mx, http://orcid.org/0000-0001-5909-1945}
\affiliation{IPICyT, Instituto Potosino de Investigaci\'on Cient\'{\i}fica y Tecnol\'ogica,\\
Camino a la presa San Jos\'e 2055, Col. Lomas 4a Secci\'on, 78216 San Luis Potos\'{\i}, S.L.P., Mexico}
\author{O. Cornejo-P\'erez}
\email{octavio.cornejo@uaq.mx}
\affiliation{Facultad de Ingenier\'{\i}a, Universidad Aut\'onoma de Quer\'etaro,\\
Centro Universitario Cerro de las Campanas, 76010 Santiago de Quer\'etaro, Mexico}
\author{S.C. Mancas}
\email{mancass@erau.edu, http://orcid.org/0000-0003-1175-6869}
\affiliation{Department of Mathematics, Embry-Riddle Aeronautical University, Daytona Beach, FL 32114-3900, USA}
%

\begin{abstract}
For the case of nonlinear second-order differential equations with a constant coefficient of the first derivative term and polynomial nonlinearities, the factorization conditions of Rosu \& Cornejo-P\'erez are approached in two ways: (i) by commuting the subindices of the factorization functions in the two factorization conditions and (ii) by leaving invariant only the first factorization condition achieved by using monomials or polynomial sequences.
For the first case the factorization brackets commute and the generated equations are only equations of Ermakov-Pinney type. The second modification is non commuting, leading to nonlinear equations with different nonlinear force terms, but the same first-order part as the initially factored equation. It is illustrated for monomials with the examples of the generalized Fisher and FitzHugh-Nagumo initial equations. A polynomial sequence example is also included.

\bs


\end{abstract}

\maketitle

\section{Introduction}
Many dynamical systems in mechanics and in physics in general are described by non linear second order differential equations or evolve under the action of internal forces with small non-linear components, especially during external forcing or along the relaxing stage after the forcing has been canceled. In their homogeneous form,
\begin{equation}\label{eq1}
  \frac{d^2x}{dt^2}+\gamma(x)\frac{dx}{dt}+f(x)=0~,
\end{equation}
these equations are traditionally known in the literature as Li\'enard equations \cite{hl-2016,csl05,iac11}, although in the case of the constant parameter $\gamma(x)=\gamma$, they may be
considered as of Duffing type, because the Duffing oscillator corresponding to $\gamma >0$ and $f(x)=r_1 x +r_2 x^3$, with $r_1$ and $r_2$ two real constants, is a representative example. The simplest physical description of (\ref{eq1}) in the latter case is that of a particle attached to a spring which provides a restoring force which is close to linear, i.e., $r_1>0$ and $|r_2|\ll 1$. Two types of springs can be introduced, known as soft and hard \cite{JS87}, for $r_2<0$ and $r_2>0$, respectively. In the case of soft springs, in the extension phase the restoring force becomes progressively weaker than for the linear spring. The hard springs which become stiffer than the linear one while increasing the extension are less frequent.

Moreover, if one goes beyond mechanical oscillators and nonlinear electronic circuits, one comes across a second important and widespread category of equations of constant $\gamma$ coefficient which are obtained by the travelling wave reduction of reaction-diffusion equations and nonlinear evolution equations. In such cases, the coefficient $\gamma$ that we denote by $\nu$ is the constant velocity of the travelling fronts \cite{Fife79}. The nonlinear force $f$ covers the phenomenology due to (bio)chemical reactions or any process capable of producing new components.

\ms

A simple way to obtain particular solutions of these non linear second order differential equations consists in using the factorization method, where
the second-order differential operator
\beq\label{oper}
D^2+\gamma(x)D+\frac{f(x)}{x}~, \qquad D=\frac{d}{dt}~,
\eeq
is factored in terms of two different first-order differential operators in the operatorial form of equation (\ref{eq1})
\beq\label{eq2}
\left(D-\phi_2(x)\right)\left(D-\phi_1(x)\right)x=0~. 
\eeq
This may provide particular solutions of (\ref{eq1}) by a single quadrature of $\left(D-\phi_1(x)\right)x=0$.
To match the factored operator in (\ref{eq2}) to the operator (\ref{oper}), the factoring functions $\phi_i$
should satisfy the conditions
\begin{align} 
&\phi_1+\phi_2+x\frac{d\phi_1}{dx}=-\gamma \label{eqc1}\\
&\phi_1\phi_2=\frac{f(x)}{x} \label{eqc2}~
\end{align}
that have been introduced in 2005 by two of the authors \cite{rcp1,rcp2}. They applied this kind of factorization to many well-known equations with polynomial nonlinearities by taking additional advantage from the polynomial factorization of the nonlinear part. The second condition shows that $\sqrt{f(x)/x}$ is the geometric mean of the functions $\phi_i$ that can be chosen
from combinations of the factors of $f(x)/x$ if $f(x)$ is a polynomial which does not have the zero degree power. This also assures that from (\ref{eq2}) one can obtain a particular solution of (\ref{eq1}) by the quadrature of $\left(D-\phi_1(x)\right)x=0$,
\beq
\int \frac{dx}{x\phi_1(x)}=t-t_0~
\eeq
since $\phi_1$ can be chosen as one of the factors of $f(x)/x$.\\

Moreover, as in supersymmetric quantum mechanics \cite{cks,dong}, the reverting of the factorization brackets has been used in \cite{rcp1,rcp2} to obtain particular solutions of equations with identical operator part, but different polynomial part $\tilde{f}$, of the form
\beq\label{eqsusy}
\frac{d^2x}{dt^2}+\gamma(x)\frac{dx}{dt}+\tilde{f}(x)=0~, \quad \tilde{f}(x)=f(x)+\phi_2(\phi_{1,x}-\phi_{2,x})x^2~,
\eeq
where the subscript $x$ denotes the derivative with respect to $x$. As well known, the reverting of the factorization brackets in quantum mechanics is equivalent to going to the Darboux-transformed partner equation of a given linear Schr\"odinger equation and it is also based on the logarithmic derivative connection between the solutions of the Riccati and Schr\"odinger equations. Such logarithmic connections between the solutions of different nonlinear evolution equations are also well known being a very useful tool for obtaining new analytic solutions \cite{Cole,LuMa}.

On the other hand, with a different grouping of terms, one can also obtain particular solutions of `supersymmetric' nonlinear equations of the form
\beq
\label{eqi}
\frac{d^2x}{dt^2}+\tilde{\gamma}\frac{dx}{dt}+f(x)=0~,  \quad \tilde{\gamma}=\gamma+(\phi_{1,x}-\phi_{2,x})x~,
\eeq
i.e., with the same polynomial nonlinearities, but a different operator part which turns nonlinear in damping. Resorting again to the mechanical and electronic circuit description, equation (\ref{eqsusy}) describes springs with additional stiffness, whereas equation (\ref{eqi}) describes more complicated oscillators that can display positive and negative damping and chaotic dynamics, such as the cases of Rayleigh's equation of violin strings and van der Pol equation of self-excited valve circuit \cite{JS87} which are amongst the simplest particular cases of (\ref{eqi}).

All these calculus properties have yielded many interesting particular solutions of the kink and soliton type for well-known nonlinear equations obtained by the traveling wave change of variables from evolution equations \cite{rcp1,rcp2,cpetal,estev07,fahmy08,oct09,yesiltas09,justin12,mr13,tiwari15,ww17,rmf17,ziem19} and have been also widely implemented in Matlab and Maple algorithms \cite{GSbook}.\\

In this article, we discuss similar nonlinear equations and their particular solutions obtained through some additional conditions and/or modifications of the factorization functions in the factorization conditions (\ref{eqc1}) and (\ref{eqc2}) for equations (\ref{eq1}) of the Duffing type (constant parameter $\gamma(x)=\gamma$) and traveling wave reductions of reaction-diffusion equations with $\gamma=\nu$.
Regarding the variable $\gamma$ class, some cases have been presented previously in \cite{rcp2} and their study with the same focus as here is left for future work.\\

In particular, we will consider here the effect of two types of modifications of the factorization brackets in equations (\ref{eqc1}) and (\ref{eqc2}):

\begin{itemize}
  \item The first modification is performed in a way that keeps invariant the two factorization conditions, which leads to a commutative factorization setting in which the reverting of the factorization brackets does not generate a new equation.
  \item The second type of modification is by adding a polynomial into the multiplication brackets in such a way that only the first factorization condition is kept invariant which generates a non-commutative factorization.
\end{itemize}
The paper is organized as follows.
In the second section, the conditions for having a commutative factorization scheme are presented together with some physical examples of this approach. In the third section, a non-commutative factorization which generalizes the Rosu and Cornejo-P\'erez factorization is introduced and some examples are presented for illustrative purposes. The conclusions are summarized in the last section.

\section{Commutative factorization setting}
We now study the consequences of interchanging the subindexes in the RCP pair of factorization conditions. This is equivalent to adding  another pair of conditions obtained 
by commuting the subindexes in both equations. However, one can instantly find that this is a minimal change since the second factorization condition keeps its form under such an interchange. Therefore, proceeding in this way, we obtain the following triplet of different factorization conditions
\begin{align}
  & \phi_1+\phi_2+x\frac{d\phi_1}{dx}= -\gamma \label{eqcc1}\\
  & \phi_2+\phi_1+x\frac{d\phi_2}{dx}=-\gamma \label{eqcc2}\\
 & \phi_2\phi_1 \left(=\phi_1\phi_2\right)=\frac{f(x)}{x}\label{eqcc3}~.
\end{align}
In this case, by comparing the first two equations, one can see that $d\phi_1/dx=d\phi_2/dx$ implying
\beq\label{r1}
\phi_2=\phi_1+c_0~,
\eeq
where $c_0$ is an arbitrary real constant. In other words, these extended (commuting) factorization conditions introduce the additional restriction on
the factoring functions of being different only by a constant. Furthermore, from (\ref{eqi}) one has $\tilde{\gamma}=\gamma$, so that the interchange of the factorization brackets does not produce a new equation in this case. Thus, in factored form, one deals with equations of the type
\beq\label{commf}
(D-\phi_1-c_0)(D-\phi_1)x=0~,
\eeq
where $\phi_1$ satisfies
\beq\label{r2}
x\frac{d\phi_1}{dx}+2\phi_1=-\gamma-c_0~,
\eeq
which is obtained by substituting (\ref{r1}) into (\ref{eqcc1}).
For constant $\gamma$, $(\ref{r2})$ implies
\beq\label{r3}
\phi_1(x)=-\frac{\gamma+c_0}{2}+\frac{\kappa_1}{x^2}~, \qquad \phi_2(x)=-\frac{\gamma-c_0}{2}+\frac{\kappa_1}{x^2}~,
\eeq
where $\kappa_1$ is an arbitrary integration constant. Besides, $f(x)$ is obtained from (\ref{eqcc3}) as
\beq\label{r4}
f(x)=\frac{\gamma^2-c_0^2}{4}x-\frac{\kappa_1\gamma}{x}+\frac{\kappa_1^2}{x^3}~.
\eeq
A direct connection, not depending on $\gamma$, between the factoring functions and the nonlinear term $f(x)$ is obtained by substituting (\ref{r1})
in (\ref{eqcc3})
\beq\label{r5}
\phi_{1,2}=\frac{\mp c_0 - \sqrt{c_0^2+4f(x)/x}}{2}~. 
\eeq

\ms

(i) {\em Case $\gamma=0$}. For this case, let us take $c_0=-2a$ and $\kappa_1=b$ in (\ref{r3}), writing the factorization functions as
\begin{equation}\label{eqg1}
   \phi_1(x)=a+\frac{b}{x^2} ~, \qquad  \phi_2(x)=-a+\frac{b}{x^2}~.
\end{equation}
These factorization functions provide the standard  Ermakov-Pinney differential equation
\begin{equation}\label{eq3}
  \frac{d^2x}{dt^2}-a^2x+\frac{b^2}{x^3}=0~,
\end{equation}
which admits the following commuting factorizations
\begin{equation}\label{eq4}
\left(D\pm a-\frac{b}{x^2}\right)\left(D\mp a-\frac{b}{x^2}\right)x=0~,
\end{equation}
providing two particular solutions from each of the first-order equations
\begin{equation} \label{eq5}
   \frac{dx}{dt}=\pm ax+\frac{b}{x}~. 
\end{equation}
For each of the signs of the linear term, these particular solutions are given by 
\begin{equation}\label{eq6}
x(t)=\pm\sqrt{-\frac{b}{a}+\frac{e^{2a(t+c)}}{a}}~, \quad x(t)=\pm\sqrt{\frac{b}{a}+\frac{e^{-2a(t-c)}}{a}}~,
\end{equation}
respectively, where $c$ is an integration constant.
These particular Ermakov solutions correspond to a different nonlinear superposition compared to that of Pinney \cite{P51}. If one writes the general Ermakov solution for $d^2x/dt^2-a^2x+b^2 x^{-3}=0$ in the known form $x_g=\sqrt{\alpha_1x_1^2+\alpha_2x_2^2+2\alpha_3 x_1x_2}$ with the superposition constants $\alpha_i$ of $x_{1,2}=e^{2ac}e^{\pm at}$ related by $\alpha_1\alpha_2-\alpha_3^2=-b^2/W^2$, where $W$ is the Wronskian determinant of $x_{1,2}$,
 then one can see that they correspond to $\alpha_1=1/a$, $\alpha_2=0$, and $\alpha_3=b/W$.

Moreover, if $\gamma=0$, one can obtain the general solution as follows. Substituting $\phi_2=\frac{f(x)}{\phi_1 x}$ in the first factorization condition, the Abel equation of the second kind \cite{estev07}
\beq\label{ab1}
x\phi_1\frac{d\phi_1}{dx}+\phi_1^2+\frac{f(x)}{x}=0~
\eeq
is obtained, which for $f(x)=-a^2x+b^2/x^3$ has the solution
\beq\label{ab2}
\phi_1(x)=\pm \sqrt{a^2+\frac{\tilde{\kappa}}{x^2}+\frac{b^2}{x^4}}~,
\eeq
where $\tilde{\kappa}$ is an integration constant. For $\tilde{\kappa}=\pm 2ab$, one obtains the previous particular cases in equation (\ref{eqg1}).
Next, from
\beq\label{ab3}
\frac{dx}{dt}=\phi_1(x)x=\pm\sqrt{a^2x^2+\tilde{\kappa}+\frac{b^2}{x^2}}
\eeq
for the positive sign, one obtains the solutions
\beq\label{ab4}
x(t)= 
\pm\frac{1}{2a}\sqrt{e^{2a(t-t_0)}-2\tilde{\kappa}
+(\tilde{\kappa}^2-4a^2b^2)e^{-2a(t-t_0)}}
\eeq
whereas for the negative sign, the solutions are
\beq\label{ab5}
x(t)=\pm\frac{1}{2a}\sqrt{e^{-2a(t-t_0)}-2\tilde{\kappa}+(\tilde{\kappa}^2-4a^2b^2)e^{2a(t-t_0)}}~,
\eeq
all of which are general Ermakov-Pinney solutions. The solutions (\ref{eq6}) are obtained for $\tilde{\kappa}=2ab$ and $t_0=-\left(c+\frac{\ln(4a)}{2a}\right)$.

\ms

(ii) {\em Case $\gamma=constant\neq 0$}. For this case, the simplest factorization is obtained by setting $\phi_1=\phi_2=\phi$ ($c_0=0$), making identical the two factorization brackets. In this special case, we have
\begin{equation}\label{eq7}
\phi(x)=-\frac{\gamma}{2}+\frac{b}{x^2}~,
\end{equation}
which one can easily verify that satisfies the triplet factorization conditions.
The obtained second order non linear differential equation is of the following Ermakov-Pinney type
\begin{equation}\label{eq8}
  \frac{d^2x}{dt^2}+\gamma\frac{dx}{dt}+\frac{\gamma^2}{4}x-\frac{\gamma b}{x}+\frac{b^2}{x^3}=0~,
\end{equation}
or in operatorial form
\begin{equation}\label{eq9}
\left(D+\frac{\gamma}{2}-\frac{b}{x^2}\right)^2x=0~
\end{equation}
which yields the particular solutions given by
\begin{equation}\label{eq10}
x(t)=\pm\sqrt{\frac{2b}{\gamma}+\frac{e^{-\gamma (t-2c)}}{\gamma}}~,
\end{equation}
where $c$ is an integration constant.

\ms

Another possible pair of factorization functions for this case is
\beq\label{eq11}
  \phi_1(x)=a_1+\frac{b}{x^2}~, \qquad  \phi_2(x)=-a_2+\frac{b}{x^2}~,
\eeq
which generate the following equation
\begin{equation}\label{eq12}
 \frac{d^2x}{dt^2}+(a_2-a_1)\frac{dx}{dt}-a_1a_2x-\frac{(a_2-a_1) b}{x}+\frac{b^2}{x^3}=0~.
\end{equation}
Thus, for $\gamma=a_2-a_1$, equation (\ref{eq12}) admits the following commuting factorizations
\begin{align}\label{eq13}
&\left(D+a_2-\frac{b}{x^2}\right)\left(D-a_1-\frac{b}{x^2}\right)x=0~,\\
&\left(D-a_1-\frac{b}{x^2}\right)\left(D+a_2-\frac{b}{x^2}\right)x=0~,
\end{align}
which lead to two particular solutions 
obtained from
\beq\label{eq14}
\frac{dx}{dt}=a_1x+\frac{b}{x}~, \qquad   \frac{dx}{dt}=-a_2x+\frac{b}{x}~.
\eeq
These solutions are
\beq\label{eq15}
 x(t)=\pm\sqrt{\frac{-b}{a_1}+\frac{e^{2a_1(t+c_1)}}{a_1}}~, \qquad  x(t)=\pm\sqrt{\frac{b}{a_2}+\frac{e^{-2a_2(t-c_2)}}{a_2}}~,
\eeq
respectively; $c_1$ and $c_2$ are integration constants.

\ms

In closing this section, we notice that multiplying each of the factorization brackets by an exponential factor in the independent variable,
\beq
e^{\pm c_0t}\left(D-\phi_1\right)e^{\pm c_0t}\left(D-\phi_1\right)x=0~,
\eeq
is another way of producing the triplet of commuting factorization conditions. However, in this case, only the constants $\pm c_0$ are introduced in the factorization brackets.

\section{Non-commutative factorization setting}

We move now to the study of additive extensions of the factorization functions,
\beq\label{tilde1}
\tilde{\phi}_1(x)=\phi_1+\epsilon_1(x)~, \qquad \tilde{\phi}_2(x)=\phi_2+\epsilon_2(x)~,
\eeq
where the $\epsilon$ functions are arbitrary functions so far.
Of course, both factorization conditions can change under the additive extension, but to keep
a link with the initial equation defined through the $\phi$ factorization functions, we are interested
in those $\tilde{\phi}$ functions for which the first factorization condition is satisfied for the same $\gamma$ parameter while
the product one is changed to a different nonlinear force $\tilde{f}$, 
\begin{align} 
& \tilde{\phi}_1+\tilde{\phi}_2+x\frac{d\tilde{\phi}_1}{dx}= -\gamma \label{tilde2}\\
& \tilde{\phi}_1\tilde{\phi}_2=\frac{\tilde{f}(x)}{x}~. \label{tilde3}
\end{align}
Therefore the factored equation $\left(D-\tilde{\phi}_2\right)\left(D-\tilde{\phi}_1\right)x=0$ is
\beq\label{tildeeq}
\frac{d^2x}{dt^2}+\gamma \frac{dx}{dt} +\tilde{f}(x)=0~.
\eeq
The additions $\epsilon_1(x)$ and $\epsilon_2(x)$ are not independent, but related through the following relation
\beq\label{epsilons}
\epsilon_1(x)=-\frac{\int \epsilon_2(x)dx}{x}~,
\eeq
obtained by substituting (\ref{tilde1}) into (\ref{tilde2}) and (\ref{tilde3}), (for zero integration constant). This condition
can be fulfilled by power functions or a finite sum of power functions.
For the monomial case, $\epsilon_1(x)=-{\rm a} x^m$ and $\epsilon_2(x)={\rm a}(m+1)x^m$, $m\in \mathbf{N}$, 
the nonlinear force $\tilde{f}(x)$ has the expression
\beq\label{tildeef}
\tilde{f}_m(x)= 
{\rm a}\big[(m+1)\phi_1-\phi_2\big]x^{m+1}-{\rm a}^2(m+1)x^{2m+1}~.
\eeq
From the physical point of view, it is useful to think of (\ref{tildeeq}) as an equation that replaces (\ref{eq1}) under small perturbations of the nonlinear force. In this perturbative context, the most interesting cases are the lowest powers, $m=0$ and $m=1$, which provide the following $\tilde{f}_m(x)$
\begin{align} \label{tildeef01}
&\tilde{f}_0(x)=f(x)+{\rm a}(\phi_1-\phi_2-{\rm a})x~, \\
&\tilde{f}_1(x)=f(x)+{\rm a}(2\phi_1-\phi_2)x^2-2{\rm a}^2x^3~.
\end{align}

\ms

\subsection{Examples}

We illustrate the monomial extension with two cases that are traveling wave frame forms of reaction-diffusion equations and also provide a finite polynomial sequence case. In the traveling wave context, the $\gamma$ parameter is the velocity $\nu$ of the traveling wave.\\

(1). {\em The generalized Fisher equation}

\ms

\begin{figure}[h!]
\centering
\subfigure[\ $n=2$; $c_1=1$, $c_2=0$.]{
\includegraphics[scale=0.780]{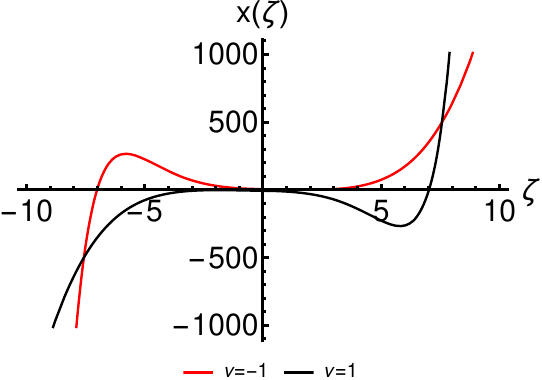}}
\subfigure[\ $n=2$; $c_1=-1$, $c_2=0$.]{
\includegraphics[scale=0.780]{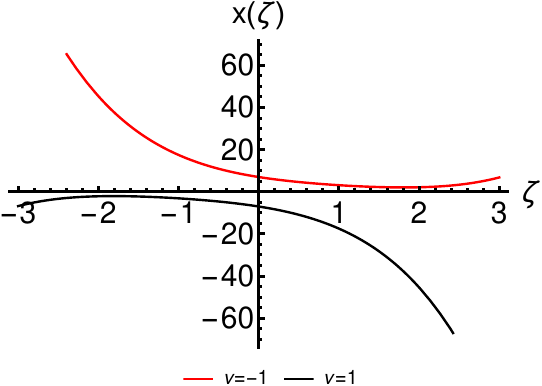}}
\caption{\label{Figs12} Particular solutions obtained from (\ref{dF1gs2}) for the values of $n$ and constants of integration as displayed.}
\end{figure}

The generalized Fisher equation has the form \cite{rcp1}
\beq\label{dF1}
x''+\nu x' +x(1-x^n)=0~,\quad \nu\neq 0~, \quad n\geq 1~,
\eeq
where the primes stand for derivatives with respect to $\zeta=s-\nu t$.
In the reaction-diffusion form, the case $n=2$ has been proposed by Fisher as an equation
governing the population dynamics in the genetics context of the alleles. It has become over the years
the fundamental law of population genetics. The general solution, obtained using {\em Mathematica},
can be written in terms of Kummer's confluent hypergeometric function of the second kind (the Tricomi function), $U$, as
\beq\label{dF1gs2}
x(\zeta)=\frac{1}{\nu}\Bigg[\frac{\zeta}{\nu}+\frac{\zeta^2}{2}- 
\frac{\zeta^{n+2}}{n+2}-U(1,n+3;\nu \zeta)+c_1 e^{\nu \zeta}\Bigg]+c_2~.
\eeq
where $c_1$ and $c_2$ are integration constants. Plots of particular solutions derived from this general Fisher solution are provided in Fig.~\ref{Figs12}.

On the other hand, equation (\ref{dF1}) can be factored with \cite{rcp1}
\beq\label{dF3}
\phi_1=h_n^{-1}\left(1-x^{n/2}\right)~, \qquad \phi_2=h_n\left(1+x^{n/2}\right)~,\qquad h_n^2=1+n/2~. 
\eeq
for $\nu_n=-\left( h_{n} + h_{n}^{-1}\right)$.
\bs

\begin{figure}[h!]
\centering
\subfigure[\ $n=2$; $\nu=-3/\sqrt{2}$, $c_0=0$.]{
\includegraphics[scale=0.780]{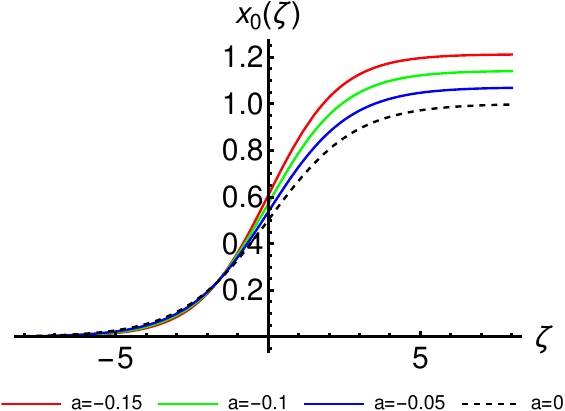}}
\subfigure[\ $n=2$; $\nu=-3/\sqrt{2}$, $c_0=0$.]{
\includegraphics[scale=0.780]{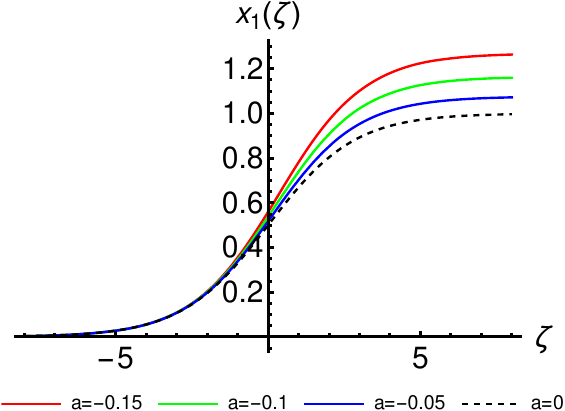}}
\caption{\label{Figs34} Particular solutions from (\ref{solpt1}) for negative values of ${\rm a}$ and the values of $n$, $\nu$, and constants of integration as displayed.}
\end{figure}

The monomially-only-extended factoring functions read
\beq\label{eqrext}
 \tilde{\phi}_1(x)=  h_n^{-1}\left(1-x^{n/2}\right)-{\rm a}x^m~, \qquad  \tilde{\phi}_2(x)= 
 h_n\left(1+x^{n/2}\right)+{\rm a}(m+1)x^m~,
\eeq
which lead to
\beq\label{tildef}
\tilde{f}(x)= f(x)+ {\rm a} \bigg[\frac{m+1}{h_n}-h_n-{\rm a} (m+1) x^m-
\left(\frac{m+1}{h_n}+h_n\right) x^{n/2} \bigg]x^{m+1}
\eeq
Using (\ref{eqrext}), a particular solution of
\beq\label{moeeq}
x''+\nu_n x' +x(1-x^n)+ {\rm a} \bigg[\frac{m+1}{h_n}-h_n-{\rm a} (m+1) x^m-
\left(\frac{m+1}{h_n}+h_n\right) x^{n/2} \bigg]x^{m+1}=0~,
\eeq
is obtained from
\beq\label{stildeeq}
\frac{dx}{d\zeta}-h_n^{-1}\left(1-x^{n/2}\right)x+{\rm a}x^{m+1}=0~,
\eeq
as
\beq\label{solpt}
\int\frac{dx}{x(h_n {\rm a}x^m-x^{n/2}+1)}=\frac{1}{h_n}\int d\zeta~.
\eeq
For $n=2$ and the cases $m=0$ and $m=1$, the quadrature in the latter equation provides the following particular solutions
\beq\label{solpt1}
x_0(\zeta)=
\frac{1-\sqrt{2}{\rm a}}{2}e^{\frac{1-\sqrt{2}{\rm a}}{2}\left(\frac{\zeta}{\sqrt{2}}-c_0\right)} 
{\rm sech} \frac{1-\sqrt{2}{\rm a}}{2}\left(\frac{\zeta}{\sqrt{2}}-c_0\right)~, 
\quad 
x_1(\zeta)=\frac{e^{\frac{\zeta}{\sqrt{2}}-c_1}}{1+\left(1+\sqrt{2}{\rm a}\right)   
e^{\frac{\zeta}{\sqrt{2}}-c_1}}~, 
\eeq
respectively, where $c_0$ and $c_1$ are integration constants. This kind of particular solutions are presented in Fig.~\ref{Figs34} and have typical traveling wave front profiles.

\ms

\begin{figure}[h!]
\centering
\includegraphics[scale=0.720]{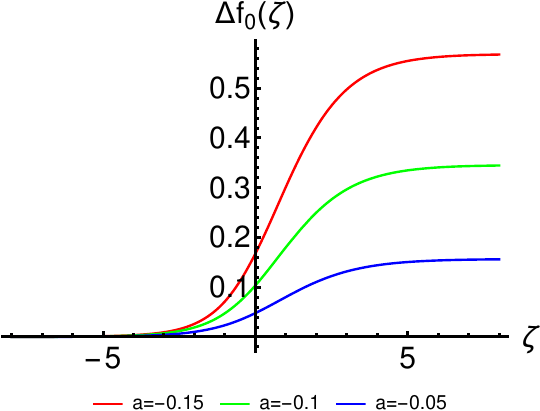} 
\includegraphics[scale=0.720]{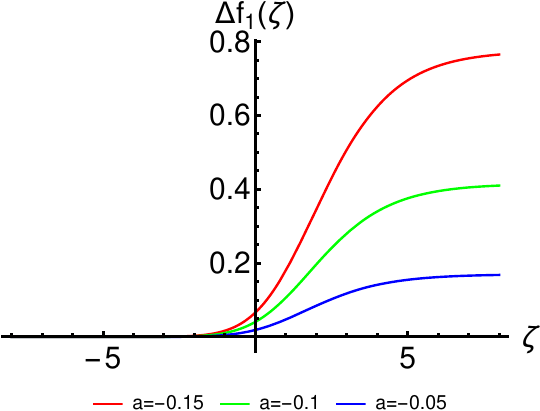} 
\caption{\label{Figs3b4b} Differences between the nonlinear forces as given by (\ref{diff-ff0}) and (\ref{diff-ff1}), respectively.}
\end{figure}

The differences between the nonlinear forces for these cases are given by the expressions
\begin{align} 
&\Delta f_0(\zeta)=\tilde{f}_0-f=-\frac{{\rm a}}{\sqrt{2}}\left(\sqrt{2}{\rm a}+1+3x_0(\zeta)\right)x_0(\zeta)~, \label{diff-ff0}\\
&\Delta f_1(\zeta)=\tilde{f}_1-f=-2{\rm a}({\rm a}+\sqrt{2})x_1^3(\zeta)~ \label{diff-ff1}
\end{align}
and are plotted in Fig.~\ref{Figs3b4b}. For small values of the parameter ${\rm a}$, they still have the switching profile of the solutions.

\bs

(2). {\em The FitzHugh-Nagumo Equation}\\

\begin{figure}[h!]
\centering
\subfigure[\ $\beta=\pm 1$; $c_1=1$, $c_2=0$.]{
\includegraphics[scale=0.780]{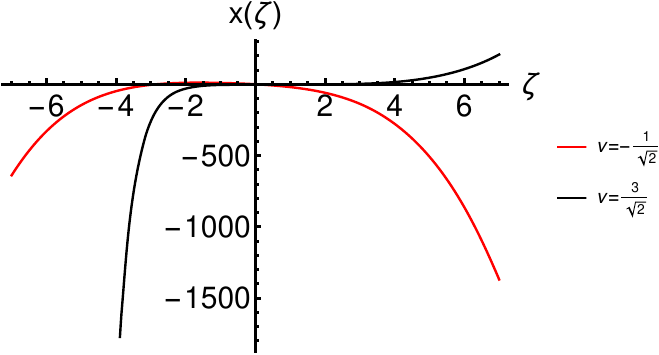}}
\subfigure[\ $\beta=\pm 1$; $c_1=-1$, $c_2=0$.]{
\includegraphics[scale=0.780]{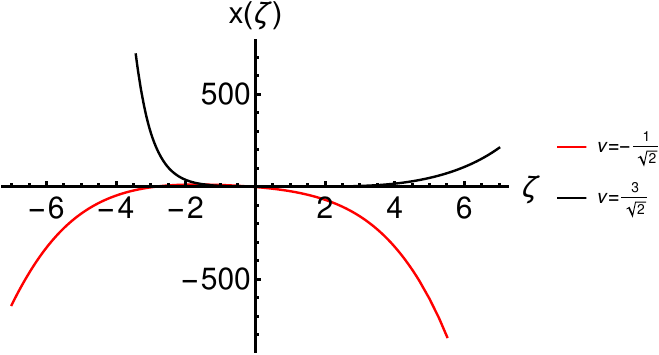}}
\caption{\label{Figs56} Particular solutions obtained from (\ref{fhngs}) for the values of $\beta$ and constants of integration as displayed.
The values of $\nu$, $-1/\sqrt{2}$ and $3/\sqrt{2}$, correspond to positive and negative $\beta$, respectively.}
\end{figure}

The FitzHugh-Nagumo equation,
\beq\label{fhn1}
x''+\nu x' + f(x)=0~, \qquad f(x)=x(x-1)(\beta-x)~,
\eeq
emerged in a simplified system of two equations modelling the transmission of electrical impulses through a nerve axon with the variable $x$ representing the axon membrane potential. In the homogeneous equation (\ref{fhn1}) the effect of a slow negative feedback on the membrane potential is not taken into account which eliminates the evolution equation of the feedback.
The general solution is
\beq\label{fhngs}
x(\zeta)=-\frac{1}{\nu}\Bigg[\frac{2p_1(\nu)+\beta \nu^2}{\nu^3}\zeta-\frac{2p_1(\nu)+\beta \nu^2}{\nu^2}\frac{\zeta^2}{2}
+\frac{p_1(\nu)}{\nu}\frac{\zeta^3}{3}-\frac{\zeta^4}{4}+\tilde{c}_1e^{-\nu \zeta}\Bigg]+\tilde{c}_2~,
\eeq
where $p_1(\nu)=3+(\beta+1) \nu$ and $\tilde{c}_{1,2}$ are arbitrary integration constants.
Some particular solutions derived from this general Fitz-Hugh-Nagumo solution are plotted in Fig.~\ref{Figs56}. Their profiles are not very different from the particular Fisher solutions obtained from the general Fisher solution.

\begin{figure}[h!]
\centering
\includegraphics[scale=0.780]{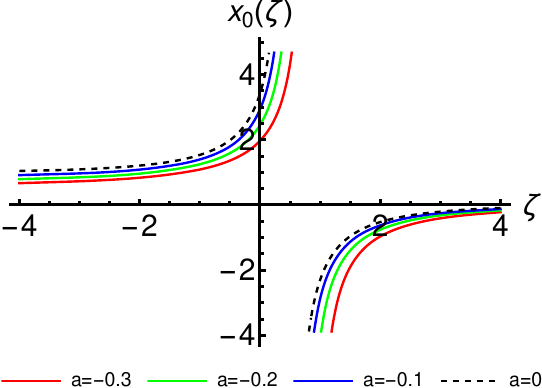} 
\includegraphics[scale=0.780]{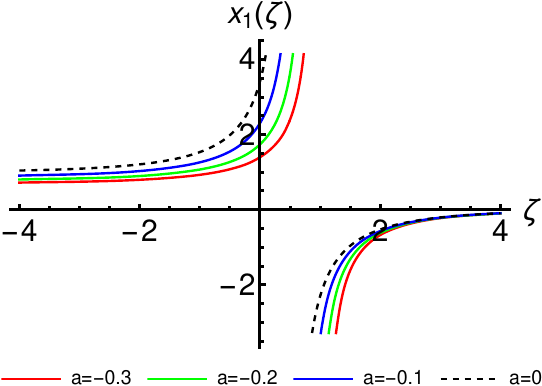} 
\caption{\label{ffhn1} Particular solutions as obtained from (\ref{fhn7}) for zero integration constants and the displayed values of the parameter ${\rm a}$.}
\end{figure}

For the particular value $\nu_\beta= (1-2\beta)/\sqrt{2}$, equation (\ref{fhn1}) is a particular case of the generalized Burgers-Huxley equation and can be factorized \cite{rcp2} with $\phi_1(x)=(x-1)/\sqrt{2}$ and $\phi_2(x)=\sqrt{2}(\beta-x)$, which we use in the monomially-only-extended factorization functions
\beq\label{fhn2}
\tilde{\phi}_1=\frac{x-1}{\sqrt{2}}-{\rm a}x^m~, \qquad \tilde{\phi}_2=\sqrt{2}(\beta-x)+{\rm a}(m+1)x^m~
\eeq
to factorize the equation
\beq\label{fhn3}
x''+\nu_\beta x' + \tilde{f}(x)=0~,
\eeq
where
\beq\label{fhn4}
\tilde{f}(x)=f(x)-{\rm a}\left(\sqrt{2}\beta+\frac{m+1}{\sqrt{2}}\right)x^{m+1}+{\rm a}\left(\sqrt{2}+\frac{m+1}{\sqrt{2}}\right)x^{m+2}-{\rm a}^2(m+1)x^{2m+1}~.
\eeq
A particular solution of (\ref{fhn3}) is obtained from
\beq\label{fhn5}
\frac{dx}{d\zeta}-\frac{1}{\sqrt{2}}(x-1)x+{\rm a}x^{m+1}=0~
\eeq
through the following quadrature
\beq\label{fhn6}
\int \frac{dx}{x(x-1-\sqrt{2}{\rm a}x^m)}=\frac{1}{\sqrt{2}}\int d\zeta~.
\eeq
For the $m=0$ and $m=1$ cases, the particular solutions are given by
\beq\label{fhn7}
x_0(\zeta)=\frac{\sqrt{2}(1+\sqrt{2}{\rm a})}{\sqrt{2}-e^{(1+\sqrt{2}{\rm a})\frac{\zeta+2c_0}{\sqrt{2}}}}~, \qquad x_1(\zeta)=\frac{\sqrt{2}}{\sqrt{2}(1-\sqrt{2}{\rm a})-e^{\frac{\zeta+2c_1}{\sqrt{2}}}}
\eeq
respectively, where $c_0$ and $c_1$ are integration constants. These solutions plotted in Fig.~\ref{ffhn1} are manifestly singular; they blow up at finite traveling variables given by $\zeta^*_0(a)=\ln 2/(\sqrt{2}+2a)-2c_0$ and $\zeta^*_1(a)=\sqrt{2}\ln[\sqrt{2}(1-\sqrt{2}a)]-2c_1$, respectively.

\ms

\begin{figure}[h!]
\centering
\includegraphics[scale=0.720]{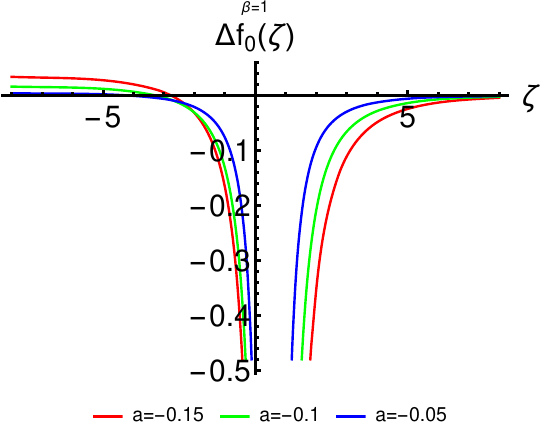} 
\includegraphics[scale=0.720]{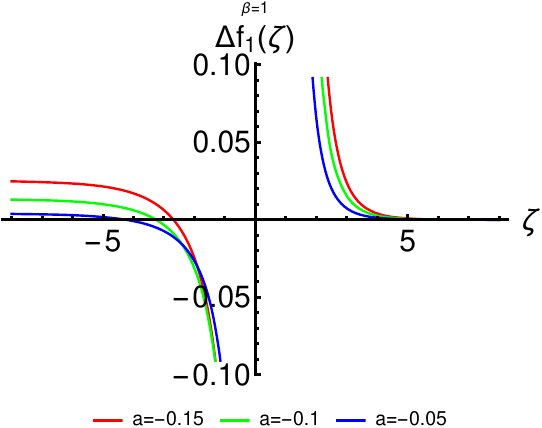} 
\caption{\label{diff-fhn} Differences between the nonlinear forces as given by (\ref{fhn8a}) and (\ref{fhn8b}), respectively, for the displayed values of the parameters.}
\end{figure}

From (\ref{fhn4}), one can also obtain the differences between the nonlinear functions of the two equations
\begin{align} 
&\Delta f_0(\zeta)=\tilde{f}_0-f=\frac{{\rm a}}{\sqrt{2}}\big[3x_0(\zeta)-2\beta-\sqrt{2}{\rm a}-1\big]x_0(\zeta)~, \label{fhn8a}\\
&\Delta f_1(\zeta)=\tilde{f}_1-f=\sqrt{2}{\rm a}\big[(2-\sqrt{2}{\rm a})x_1(\zeta)-\beta-1\big]x_1^2(\zeta)~, \label{fhn8b}
\end{align}
for $m=0$ and $m=1$, respectively. These differences are plotted in Fig.~\ref{diff-fhn} for several values of the parameter ${\rm a}$ and $\beta=1$.

The interesting feature to be noticed is that the particular solutions obtained for the monomially-only-extended FitzHugh-Nagumo equation depend only on the parameter ${\rm a}$, while the forces depend also on the parameter $\beta$. This is due to the fact that the factoring functions $\phi_1(x)$ and $\tilde{\phi}_1(x)$ do not depend on $\beta$.

\bs

(3). {\em Polynomial sequence example}\\

Finally, we discuss a polynomial sequence extension of the factorization functions involving $N$ terms of a $\gamma=0$ initial case for
which the factorization functions $\phi_{1,2}$ are both zero, i.e., a degenerate $D^2x=0$ case.
Then, we have
\beq\label{eq16}
   \tilde{\phi}_1(x)=-\sum_{m=0}^{N}{\rm a}_mx^m~, \qquad  \tilde{\phi}_2(x)=\sum_{m=0}^{N}(m+1){\rm a}_mx^m~,
\eeq
which one can easily verify that satisfies the conditions given in equations (\ref{tilde2}) and (\ref{tilde3}).

\ms

Using the first pair of factorization functions, 
the corresponding second order non linear differential equation has the form
\beq\label{eq17}
  \frac{d^2x}{dt^2}-\left(\sum_{m=0}^{N}{\rm a}_mx^m\right)\left(\sum_{m=0}^{N}(m+1){\rm a}_mx^m\right)x \equiv
  \frac{d^2x}{dt^2}-\Bigg[\sum_{m=0}^{N}\left(\sum_{l=0}^{m}(m-l+1){\rm a}_l{\rm a}_{m-l}\right)x^m\Bigg]x=0~,
\eeq
which admits the following non-commuting factorization
\begin{equation}\label{eq18}
\left(D-\sum_{m=0}^{N}(m+1){\rm a}_mx^m\right)\left(D+\sum_{m=0}^{N}{\rm a}_mx^m\right)x=0~.\\
\end{equation}
One particular solution for equation (\ref{eq17}) can be obtained from the first order equation
\begin{equation}\label{eq19}
\frac{dx}{dt}=-\sum_{m=0}^{N}{\rm a}_mx^{m+1}~. \\
\end{equation}
While for $N<2$ one can easily obtain simple explicit solutions of (\ref{eq19}), for $N\geq 2$, the solutions will be in general implicit solutions depending on the roots of the cubic, quartic, a.s.o., algebraic equations.

\ms

Let us consider the $N=2$ case for which $\sum_{m=0}^{2}{\rm a}_mx^{m+1}=x({\rm a}_0+{\rm a}_1x+{\rm a}_2x^2)={\rm a}_2x(x-\alpha_1)(x-\alpha_2)$, where $\alpha_{1,2}$ are the roots of the quadratic algebraic equation. Then, we have the quadrature
\begin{equation}\label{eqexf1}
\int \frac{dx}{x(x-\alpha_1)(x-\alpha_2)}=-{\rm a}_2\int dt~.
\end{equation}
The classification of the solutions in terms of the roots is the following \cite{mr2016}:

\begin{itemize}
\item [(i).] If $\alpha_{1,2}=\frac{1}{2{\rm a}_2}\left(-{\rm a}_1\pm\sqrt{\Delta}\right)~, \quad \Delta={\rm a}_1^2-4{\rm a}_0{\rm a}_2 > 0$,
then by the method of partial fraction decompositions, one obtains
\begin{equation}\label{eqexf3}
\frac{1}{\alpha_1\alpha_2(\alpha_1-\alpha_2)}\bigg[\ln x^{(\alpha_1-\alpha_2)} +\ln(x-\alpha_1)^{\alpha_2}-\ln(x-\alpha_2)^{\alpha_1}\bigg]=-{\rm a}_2(t-t_0)~,
\end{equation}
which leads to the implicit solution
\begin{equation}\label{eqexf3b}
\frac{(x-\alpha_1)^{\alpha_2}}{(x-\alpha_2)^{\alpha_1}}x^{(\alpha_1-\alpha_2)}=e^{-{\rm a}_2\alpha_1\alpha_2(\alpha_1-\alpha_2)(t-t_0)}\equiv
e^{-\frac{{\rm a}_0}{{\rm a}_2}\sqrt{\Delta}(t-t_0)}~.
\end{equation}

\ms

\item [(ii).] When $\alpha_1=\alpha_2=\alpha\,\, (=-\frac{{\rm a}_1}{2{\rm a}_2})$, $\Delta=0$, the implicit solution is
\beq\label{case2}
-\frac{1}{x-\alpha} + \frac{1}{\alpha}\ln\bigg|\frac{x}{x-\alpha}\bigg|=-\alpha\left({\rm a}_2t+{\rm c}\right)\equiv \frac{{\rm a}_1}{2}(t-t_0)~.
\eeq

\ms

\item [(iii).] If $\alpha_1= \bar{\alpha}_2=r+ i s$, $\Delta<0$, the implicit solution is
\beq\label{case3}
-\ln\big|\sqrt{(x-r)^2+s^2}\big|+\frac{\ln |x|}{s}+\frac{r}{s}\arctan\frac{x-r}{s} 
={\rm a}_2(r^2+s^2)(t-t_0)~. 
\eeq

\item [(iv).] In the degenerate case $\alpha_1=\alpha_2=0$, i.e., ${\rm a}_0={\rm a}_1=0$, one obtains the simple explicit solution
\beq\label{case4}
\frac{1}{2x^2}={\rm a}_2(t-t_0)~. 
\eeq

\end{itemize}
\noindent In all cases, $t_0$ is an arbitrary integration constant.

\ms

In a very limited amount of these kinds of zero $\gamma$ cases, one can also obtain implicit solutions with two integration constants (general solutions). As in the Ermakov-Pinney case, this is obtained through the Abel equation of the second kind for the factorization function $\tilde{\phi}_1$,
which reads
\beq\label{ab2nd1}
\tilde{\phi}_1\frac{d\tilde{\phi}_1}{dx}+\frac{1}{x}\tilde{\phi}_1^2=\sum_{m=0}^{N}\left(\sum_{l=0}^{m}(m-l+1){\rm a}_l{\rm a}_{m-l}\right)x^{m-1}~.
\eeq
In terms of the function $\psi= \tilde{\phi}_1^2$, this equation is a linear first order equation, which in the $N=2$ leads to the solution
\beq\label{ab2nd2}
\tilde{\phi}_1(x)=\pm \sqrt{({\rm a}_0+{\rm a}_1x+{\rm a}_2x^2)^2+\frac{k_1}{x^2}}~,
\eeq
where $k_1$ is an integration constant. Then, from
\beq\label{ab2nd3}
\frac{dx}{dt}=\tilde{\phi}_1(x)x=\pm\sqrt{({\rm a}_0x+{\rm a}_1x^2+{\rm a}_2x^3)^2+k_1}~,
\eeq
one can obtain for ${\rm a}_0={\rm a}_1=0$, ${\rm a}_2\neq 0$ (case (iv) above) the general implicit solution
\beq\label{ab2ndd4}
\sqrt{(a_2x^3)^2+k_1}\,{}_{2}F_{1}\left(\frac{2}{3}, 1, \frac{7}{6}; -\frac{(a_2x^3)^2}{k_1}\right)=k_1(t-t_0)~,
\eeq
where 
$_{2}F_{1}$ is Gauss' hypergeometric function. 
For $k_1=0$, one obtains the explicit singular solution given in (\ref{case4}).

\bs

\section{Conclusions}

We have discussed some minimal extensions of the factorization conditions of Rosu and Cornejo-P\'erez in the case of the constant $\gamma$ coefficient of the first derivative with emphasis on the generated nonlinear equations and their particular solutions. The necessary conditions to have commutative factorizations have been introduced which lead to equations of the Ermakov-Pinney type at most as has been shown in this paper. For the non-commutative factorization case, one can obtain equations with the same $\gamma$ parameter through designed additive monomial extensions of the factorization functions. The new equations have nonlinear forces that differ from the initial nonlinear elastic forces by supplementary terms. In the mechanical context of spring models, one may seek applications in cases of weak nonlinear nanoelasticity \cite{ne1,ne2}. On the other hand, in this paper, the illustrative examples have been chosen from the vast area of reaction-diffusion equations in which a huge variety of travelling fronts are present and $\gamma$ is just the constant velocity of their motion. We have presented such kinds of modified counterparts of the generalized Fisher equation and the FitzHugh-Nagumo equation, and their particular solutions have been obtained by the factorization method. A polynomial sequence extension in the case $\gamma=0$ has been also provided, for which various types of implicit solutions have been given for the $N=2$ size of the sequence.
We hope to extend these ideas in future work to the more general case of dissipation depending on the spatial coordinate and time and also to inhomogeneous equations of this kind \cite{inhom} with the goal of enlarging the class of nonlinear equations with analytic solutions and identifying their possible applications.

\bs

\subsection*{CRediT authorship contribution statement}

\noindent {\bf G. Gonz\'alez}: Conceptualization, Methodology, Writing - review \& editing.

\noindent {\bf H.C. Rosu}: Writing - original draft, Supervision, Formal analysis.

\noindent {\bf O. Cornejo-P\'erez}: Investigation, Conceptualization, Writing - review.

\noindent {\bf S.C. Mancas}: Calculations, Investigation, Writing - review \& editing.

\subsection*{Declaration of competing interests}

The authors declare they have no known competing financial interests or personal relationships that could
have appeared to influence the work reported in this paper.


\end{document}